\documentclass[aps,prc,tightenlines,showpacs,superscriptaddress,twocolumn]
{revtex4}

\usepackage{graphicx,color}

\begin{document}

\title{The Neutron Star Inner Crust: Nuclear Physics Input}

\author{Andrew W. Steiner} 
\affiliation{Joint Institute for Nuclear Astrophysics, National
Superconducting Cyclotron Laboratory, and the Department of Physics
and Astronomy, Michigan State University, East Lansing, MI 48824}

\begin{abstract} 
A fully self-consistent model of the neutron star inner crust based
upon models of the nucleonic equation of state at zero temperature is
constructed. The results nearly match those of previous calculations
of the inner crust given the same input equation of state. The extent
to which the uncertainties in the symmetry energy, the
compressibility, and the equation of state of low-density neutron
matter affect the composition of the crust are examined. The
composition and pressure of the crust is sensitive to the description
of low-density neutron matter and the nuclear symmetry energy, and the
latter dependence is non-monotonic, giving larger nuclei for moderate
symmetry energies and smaller nuclei for more extreme symmetry
energies. Future nuclear experiments may help constrain the crust and
future astrophysical observations may constrain the nuclear physics
input.
\end{abstract}


\maketitle

\section{Introduction}

The inner crust of a cold neutron star can be defined as the region
between the density where neutrons drip out of nuclei (about 4 $\times
10^{11}$ g/cm$^{3}$) and the density for the transition to homogeneous
nucleonic matter at about half of the nuclear saturation density. This
region is sensitive to the nuclear physics input because the
nature of the crust is determined by the structure of neutron-rich
nuclei and the energetics of the surrounding dripped neutrons. In
this work, the dependence of the description of the 
neutron-rich nuclei and the dripped neutrons on the equation of state
(EOS) of homogenous nucleonic matter is examined.

The inner crust is of broad interest because a large variety of
astrophysical observations are dependent on and sensitive to the
properties of the neutron star crust. One recent motivation is the
suggestion that the giant flares in Soft Gamma-Ray repeaters trigger
seismic events in the neutron star crust and are sensitive to the
shear modulus of the crust
crust~\cite{Thompson95,Strohmayer06,Samuelsson07}. The shear modulus,
in turn, is sensitive to the composition of the neutron star crust and
the relative magnitude of the proton and neutron numbers of the nuclei
in the inner crust. The moment of inertia of rotating neutron stars is
also sensitive to the inner crust and depends on the transition
density between the crust and the core~\cite{Link99}.  Neutrino and
photon opacities are also sensitive to the properties of the nuclei in
the inner crust. For example, neutrino-nucleus scattering, which
scales like $A^2$, is the most important neutrino process during the
lepton-trapped phase of a Type II supernova (see Ref.~\cite{Burrows06}
for a recent review). Finally, the cooling and evolution of neutron
star crusts depends on the both the size of the
crust~\cite{Lattimer94} and by its transport properties
\cite{Potekhin97,Brown00}, which are both related to the
composition. These astrophysical connections motivate the study of the
magnitude of the uncertainty of the properties of the inner crust
which come from present uncertainties in the nuclear physics inputs.

In this article, several models of the neutron star inner crust are
constructed systematically using inputs from with the current
experimental information while allowing the range of uncertainty
allowed due to the uncertainty in the EOS of homogeneous matter. Of
particular importance, is that the symmetry energy is varied in {\em
both} the description of the nuclei and the description of the neutron
matter at the same time. The composition depends on the
symmetry energy, but is nearly independent of the
compressibility. This means that astrophysical observations which are
connected to properties in the crust can constrain the nuclear
symmetry energy.

The inner crust is quite sensitive to the EOS of neutron matter at
sub-saturation densities. At sufficiently low densities, neutron
matter is somewhat well understood because three-body interactions are
small, and the two-body neutron-neutron interaction is strongly
constrained by the experimentally measured neutron-neutron scattering
phase shifts~\cite{Carlson03} (see also the review in
Ref.~\cite{Heiselberg00}). Many of the currently available EOSs,
however, do not respect this understanding of low-density neutron
matter because they are fit to the properties of nuclei which are more
sensitive to matter near saturation densities.  The neutron matter
EOSs used in this work are designed to have a realistic behavior below
the saturation density, within the precision required for the
description of the crust.

\section{The Mass Models}

While microscopically-based models of the nuclei are of great interest
because they can disentagle important effects which are not easily
treated in a classical approach, a microscopic approach can also make
it more difficult to understand the physical principles which guide
the nature of the inner neutron star crust. In addition, it is not
clear that a classical approach is significantly less effective at
estimating the magnitude of uncertainties originating in the nuclear
physics input (it may even be {\it more} effective). In any case,
since the purpose is only to estimate the uncertainties from the
nuclear physics input to the EOS, a liquid-drop model quite similar to
that described in Refs.~\cite{Baym71b,Lattimer85} is used. More
microscopic models for the crust have been developed (see the
pioneering work of Ref.~\cite{Negele83} and recent efforts in
Refs.~\cite{Baldo07,Newton07}) and it is expected that these results on the
sensitivity to the EOS of homogeneous nucleonic matter will apply to
some extent in these models as well.

The liquid-drop model for this work consists of
a bulk energy contribution which is determined from the EOS of
homogenous nucleonic matter together with surface and Coulomb
contributions. This will be compared to the finite-range droplet model
described in~Ref.~\cite{Myers69} and used in Ref.~\cite{Moller95}.

The binding energy per baryon of a nucleus with proton number $Z$ and
atomic number $A$ is given by 
\begin{eqnarray}
B(Z,A)/A &=& B_{\mathrm{bulk}}(n_n,n_p)/A + 
\sigma {\cal B}(n_n,n_p) 
\left(\frac{36 \pi}{n^2 A}\right)^{1/3} \nonumber \\
&& + {\cal C}~\varepsilon_{\mathrm{Coulomb}}/n
\label{eq:massform}
\end{eqnarray}
where $n_n$ and $n_p$ are the average neutron and proton densities 
inside the nucleus with the given $Z$ and $A$. The binding energy 
of bulk matter, $B_{\mathrm{bulk}}$
(about $-$16 MeV in isospin symmetric matter) is given by
\begin{equation}
B_{\mathrm{bulk}}= \frac{A}{n}
\left[\varepsilon(n_n,n_p) - n_n m_n - n_p m_p\right]
\end{equation}
where $m_n$ and $m_p$ are the neutron and proton masses (which is
taken to be 939 MeV), $n=n_n+n_p$ is the average baryon number density
in the nucleus, and $\varepsilon(n_n,n_p)$ is the energy density of
homogeneous matter evaluated at the given neutron and proton
density. The expression for $\varepsilon$ may be given by any EOS of
homogeneous matter and several different models are employed. Note
that the energy of the dripped neutrons which is added later will
always be determined with the same EOS as is used to describe the bulk
part of the nuclear energy.

The average baryon density will be determined from
\begin{equation}
n = n_n+n_p = n_0 + n_1 I^2
\label{eq:n}
\end{equation}
where $I = 1 - 2 Z/A$. The parameter $n_0$ is analogous to the
saturation density of nuclear matter and is expected to be near 0.16
fm$^{-3}$. The parameter $n_1$ subsumes (in a very schematic way) two
effects: the decrease in the saturation density with the isospin
asymmetry and the increase in the saturation density due to the
Coulomb interaction. These effects are both explicitly present in the
finite range droplet model (see Eq. 49 of Ref.~\cite{Moller95}). The
decrease in the saturation density with isospin asymmetry is typically
larger and thus $n_1$ is always negative in these models.

The individual average neutron and proton
number densities are given by
\begin{eqnarray}
n_n & = & n (1 + \delta) / 2 \nonumber \\
n_p & = & n (1 - \delta) / 2
\label{eq:np}
\end{eqnarray}
and the density asymmetry $\delta= 1 - 2 n_p / (n_n + n_p) $ is given
by $\delta = \zeta I $ where $\zeta$ is a constant parameter of the
model.  Neutron and proton radii (``squared-off'' radii, not
root-mean-square radii) are given simply by $4 \pi n_n R_n^3 = 3N$,
and $4 \pi n_p R_p^3 = 3Z$. The presence of a neutron skin is
determined from $\zeta$. If $\zeta$ is unity, then all nuclei have no
neutron skin ($R_n=R_p$), while if $\zeta$ is less than unity, then
all nuclei with $N>Z$ will have a neutron skin ($R_n>R_p$).

The surface energy contribution is proportional to the surface tension 
$\sigma$, $A^{2/3}$ (the surface energy scales as $A^{2/3}$ so that
the surface energy per baryon scales like $A^{-1/3}$ as in
Eq.~\ref{eq:massform}), and a unitless function ${\cal B}$. 
Typically this latter function is quadratic in the isospin symmetry 
\begin{equation}
{\cal B}(n_n,n_p) = 1 - {\sigma}_{\delta} \delta^2
\end{equation}
where $\sigma_{\delta}$ is a positive parameter representing the
surface symmetry energy. This is essentially the approach taken in
Ref.~\cite{Myers69}.  For a neutron star inner crust model, this can
be modified to ensure that the surface energy vanishes in the limit
$\delta \rightarrow 1$ as it must. One possible approach (and the one
used here) is that from Ref.~\cite{Lattimer85}
\begin{equation}
{\cal B}(n_n,n_p) = \frac{16 + b}{\left[1/x^3 + b + 1/(1-x)^3 \right]}
\end{equation}
where $x=n_p/n$ and $b$ is a simple function of the parameter
$\sigma_{\delta}$ and is related through $\sigma_{\delta}=96 \sigma
/(b+16)$. This is an approximate scheme for taking into account the
isospin properties of the surface energy which may suffice for the
present purpose, but note the more the more detailed discussion
in Ref.~\cite{Steiner05}. In particular there is still an unresolved
ambiguity associated with how the surface energy is handled as
discussed in this reference. The slope of the correlation between the
surface symmetry energy, $\sigma_{\delta}$ and the symmetry energy at
the saturation density depends on the mass formula used. This model,
like all other present models for the neutron star crust, effectively
chooses a particular slope for this correlation.

The Coulomb energy density of a 3-dimensional droplet of protons can
be written~\cite{Baym71b,Ravenhall83} (modulo an overall factor of
$\chi$, the volume fraction of matter present in nuclei, which is
included later),
\begin{equation}
\varepsilon_{\mathrm{Coulomb}} 
= \frac{2 \pi}{5} n_p^2 e^2 R_p^2 \left( 2-3 \chi^{1/3}+\chi \right)
\end{equation}
where $e^2$ is the usual Coulomb coupling~$\sim \hbar c/137$.  In the
final term in parenthesis, the first term corresponds to the standard
Coulomb contribution, the second term corresponds to the ``lattice
contribution''~\cite{Baym71} in the Wigner-Seitz approximation, and
the last term to a further finite-size correction relevant at higher
densities when $\chi$ is comparable to unity.  Note that this last
term is quite important near the crust-core transition and tends to
delay the transition to nuclear matter to higher densities. 
The Coulomb contribution is multiplied by a parameter ${\cal C}$ to take
into account the fact that the proton density does not fall off
sharply at a finite radius and this surface diffusiveness maybe
dependent on the input symmetry energy. This parameter will always be
nearly unity. As noted in Ref.~\cite{Chamel07}, the Wigner-Seitz
approximation fails when describing the low-temperature transport
properties, but will suffice for describing the composition and the
equation of state as done here.

In summary, there are six free parameters in this model (outside of
the input equation of state of bulk nuclear matter, which is a kind of
parameter in itself) are the surface tension in MeV/fm$^{2}$,
$\sigma$, the surface symmetry energy $\sigma_{\delta}$, the
correction factor to the Coulomb energy, $ {\cal C} $, the asymmetry
parameter $\zeta$, and the central density parameters, $n_0$ and
$n_1$ which are expressed in units of fm$^{-3}$. 
These six parameters will be fit
to experimental masses for each input EOS.

\section{The Equations of State}

The EOS from Ref.~\cite{Akmal98} (APR) is used, which was obtained
from variational chain summation calculations of the equation of state
using a realistic nucleon-nucleon interaction. Also, a ``typical''
relativstic field-theoretical model is utilized (review in
Ref.~\cite{Serot89}), NL4~\cite{Nerlo-Pomorska04} which was fit to
nuclei. In order to compare with the model of Ref.~\cite{Douchin01} 
the Skyrme~\cite{Skyrme59} model SLy4~\cite{Chabanat95} is used, and
in order to compare with the model from Ref.~\cite{Lattimer06} 
the Skyrme model SkM$^{*}$~\cite{Bartel82} is used.

APR is expected to be particularly good for neutron matter at low
densities, because it is directly computed from an interaction which
reproduces the two-body nucleon-nucleon phase shifts. The model SLy4
also has a good neutron matter EOS because it was fit to both nuclei
and low-density neutron matter. The NL4 and SkM$^{*}$ models were only
fit to nuclei and low-density neutron matter are less
constrained. SkM$^{*}$ happens to have a neutron matter EOS which is
somewhat closer to APR than NL4. Like the SLy4 interaction,
relativistic models are also able to reproduce, at some level, the
more accurate low-density neutron matter EOS found in APR and SLy4, as
was demonstrated by the RAPR model in Ref.~\cite{Steiner05} and the
FSUGold model~\cite{Todd-Rutel05,Piekarewicz07}.

In order to examine the importance of having an accurate EOS for
low-density neutron matter, the low-density neutron matter EOS of NL4
is modified and compared to the original.  Three new models are
constructed. The first model, NL4Q, is a modification of NL4 which
treats the symmetry energy at low densities to be exactly
quadratic. This approximation is quite good at lower densities, and
the NL4 crust is nearly indistinguishable from the NL4Q crust.  The
other models, NL4QN and NL4QN2, are versions of NL4Q which reproduces
the neutron matter EOS of APR at densities below a specified density,
\begin{equation}
E_{\mathrm{neut}}^{\mathrm{NL4QN}}=E_{\mathrm{neut}}^{\mathrm{APR}}+
\frac{ E_{\mathrm{neut}}^{\mathrm{NL4Q}}-
E_{\mathrm{neut}}^{\mathrm{APR}} }
{ 1+ e^{(n_t-n)/\nu} }`
\end{equation}
where $n_t$ is 0.08 (0.04) fm$^{-3}$ and $\nu$ is 0.0105 (0.016)
fm$^{-3}$ for model NL4QN (NL4QN2). Both NL4QN and NL4QN2 have
pressures and neutron chemical potentials which monotonically increase
with density. The results below show that these two models give
signficantly different results for the inner crust.

Finally, several schematic models of the EOS are constructed so that
the effect of the compressibility and the symmetry energy can be
examined. The schematic EOS for neutron matter is
\begin{eqnarray}
E_{\mathrm{neut}}/A &=& \left[1-0.6~k_{F,n}^{0.4} + \eta_1 \left(\frac{n}{n_0}\right) + 
\eta_2 \left(\frac{n}{n_0}\right)^2 \right] \nonumber \\
&& \times \frac{k_{F,n}^5}{10 \pi^2 m_n}
\label{eq:neut}
\end{eqnarray}
where $k_F = (3 \pi^2 n)^{1/3}$ is the neutron Fermi momentum and the
last term is just the free Fermi gas energy density. The first two terms
inside the square brackets are designed to reproduce the expectation
from equations of state at low densities obtained from two-body
potentials which reproduce the experimental phase-shift data on
neutron-neutron scattering~\cite{Carlson03,Gezerlis07}.  It is
expected that the interacting neutron matter EOS is
about half the free Fermi gas energy at $k_{F,n}=0.5$ fm$^{-1}$. The term
proportional to $k_{F,n}^{0.4}$, qualitatively reproduces this low-density
behavior and the other parameters $\eta_1$ and $\eta_2$ can be
adjusted for densities near the saturation density where the EOS is
more uncertain. 

\begin{table}
\begin{tabular}{ccccc}
Model & $\eta_1$ (MeV) & $\eta_2$ (MeV) 
& $E_{\mathrm{sym}} (MeV)$ & $\gamma$ \\
\hline
Sch          & -0.307 & 0.481 & 31 & 0.9 \\
SchS28       & -0.487 & 0.578 & 28 & 0.9 \\
SchS34       & -0.127 & 0.385 & 34 & 0.9 \\
Sch$\gamma$1 & -0.0308 & 0.198 & 31 & 0.6 \\
Sch$\gamma$2 & -0.793 & 0.979 & 31 & 1.1 \\
\hline
\end{tabular}
\caption{The values of $\eta_1$ and $\eta_2$ (c.f. Eq.~\ref{eq:neut}),
parameters controlling the neutron matter equation of state in the
schematic models and the corresponding values of the symmetry energy
at saturation density and the exponent $\gamma$. In the schematic
models, $\eta_0$ is fixed at 0.5. }
\end{table}

It is useful to connect this description with the more traditional
description neutron matter in terms of a symmetry energy with the form
$ E_{\mathrm{sym}} = A (n/n_0)^{2/3} + B (n/n_0)^{\gamma}$. For an
effective mass of about 0.7 $M$, $A$ is about 17 MeV, and then $B$ and
$\gamma$ dictate the magnitude of the symmetry energy at the
saturation density and the density dependence of the symmetry energy,
respectively. The base model schematic model, ``Sch'', has values of
$\eta_1$ and $\eta_2$ appropriate for $B=14$ MeV (a symmetry energy of
31 MeV) and $\gamma=0.9$. These values are given in Table~I, as well
as the values of $\eta_1$ and $\eta_2$ for the schematic models whose
symmetry energy is different from the baseline model. The variation of
the value of the symmetry energy at saturation density between 28 and
34 MeV is consistent with the observation that most modern equations
of state fall within this range. When the symmetry energy is taken to
be a pure power law (the A=0 limit), the limit of the variation of
$\gamma$ ($0.6<\gamma<1.1$) is inferred from the experimental
information from intermediate-energy heavy-ion
collisions~\cite{Tsang04,Chen05}. The models Sch$\gamma$1 and
Sch$\gamma$2 are constructed by fitting a symmetry energy with the
given exponent with A=0 to express the expected range. The extraction
of values of $\gamma$ from heavy-ion collisions is non-trivial, and
there may be systematic uncertainties that are not yet
understood. These uncertainties would mean that the range of variation
presented here is overly conservative, and that the true range might
be larger.

The schematic equation of state for nuclear matter is
\begin{eqnarray}
E_{\mathrm{nuc}}/A &=& M + B + \frac{K}{18 n_0^2} \left(n - n_0\right)^2
\nonumber \\
&& + \frac{K^{\prime}}{162 n_0^3} \left(n - n_0\right)^3
\end{eqnarray}
where $M$ is the nucleon mass, $B$ is the binding energy, $n_0$ is the
saturation density, $K$ is the compressibility and $K^{\prime}$ is the
``skewness''.  Isospin asymmetric matter is computed assuming that the
symmetry energy is exactly quadratic in the isospin asymmetry $\delta$
(this approximation may fail at high density, see
Ref.~\cite{Steiner06}).  Note that varying the compressibility in this
model is not precisely equal to varying the quantity which might be
obtained from giant resonances, as the latter are only sensitive to
the equation of state in the neighborhood of saturation density
whereas the compressibility is applied to nuclear matter at all
densities below the saturation density, thus the variation in the
compressibility is a bit larger than that recently suggested in
Refs.~\cite{Agrawal03,Colo04,Garg06}. The baseline schmatic model
``Sch'', has a binding energy of $-$16 MeV, a saturation density of
0.16 fm$^{-3}$, a symmetry energy of 31 MeV, a compressibility of 230
MeV. In all of the models, the skewness parameter is fixed by ensuring
that the energy per baryon of nuclear matter vanishes at zero density,
as it ought. In addition to variations of the symmetry energy as
discussed above, Two models ``SchK210'' and ``SchK250'', are
constructed to be the same as the baseline model, except that they
have different compressibilities.

\begin{figure}
\includegraphics[scale=0.42]{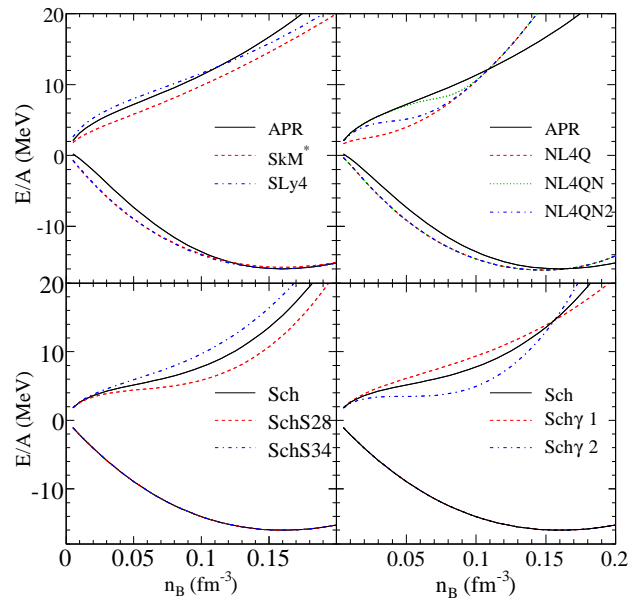}
\caption{A survey of the equations of state used in this work. Plotted
in each panel are the energy per baryon of nuclear matter (lower set
of curves) and neutron matter (upper set of curves) as a function of
baryon density, $n_{B}$. The upper left panel shows the models APR,
SLy4, and SkM$^{*}$, the upper right panel shows the model NL4Q (based
on NL4) and shows how neutron matter was modified to match APR at low
densities, and the lower panels show how the symmetry energy was
modified in the schematic equation of state.}
\label{fig:eos}
\end{figure}

A survey of some of the equations of state is given in
Fig.~\ref{fig:eos}. The upper right panel shows how neutron matter in
the NL4QN and NL4QN2 models interpolates between APR at low densities
and the normal NL4Q model at higher densities, while leaving the
nuclear matter EOS unmodified. 

\section{The Mass Fits}

The liquid drop model is fit to the experimental nuclear masses from
Ref.~\cite{Audi03} using the fitting formula
\begin{equation}
\Delta_{\mathrm{RMS}}= \left[ \frac{1}{N} \sum_{i=1}^{N} 
\left(M^{exp}_i-M^{th}_i\right)^2 \right]^{1/2}
\end{equation}
where $N$ is the number of nuclei, and $M^{exp}$ and $M^{th}$ are the
experimental and theoretical values of the mass excess.
Ref.~\cite{Moller95} points out that this fitting formula can be
improved and that it overestimates the actual model error, but these
considerations will not be important at the level of the results 
presented here. The fitting results are given in Table~II. The fitting
results for NL4QN and NL4QN2 are not given because they were found to
be nearly equal to those from NL4Q.

\begin{table}
\label{tab:fit}
\begin{tabular}{cccccccc}
Model & $\zeta$ & $\sigma$ & $\sigma_{\delta}$ & 
${\cal C}$ & $n_1$ & $n_0$ & $\Delta_{\mathrm{RMS}}$ \\
\hline
APR & 0.886 & 1.19 & 1.72 & 0.885 & $-$0.128 & 0.181 & 2.61 \\
SkM$^{*}$ & 0.888 & 1.14 & 1.16 & 0.899 & $-$0.0612 & 0.17 & 2.61 \\
SLy4 & 0.885 & 1.19 & 1.57 & 0.882 & $-$0.11 & 0.181 & 2.6 \\
NL4Q & 0.89 & 1.15 & 2.67 & 0.915 & $-$0.234 & 0.169 & 2.66 \\
Sch & 0.897 & 1.19 & 1.67 & 0.903 & $-$0.12 & 0.176 & 2.7 \\
SchK210 & 0.897 & 1.19 & 1.69 & 0.907 & $-$0.117 & 0.174 & 2.72 \\
SchK250 & 0.897 & 1.19 & 1.64 & 0.9 & $-$0.12 & 0.176 & 2.69 \\
SchS28 & 0.891 & 1.2 & 1.18 & 0.892 & $-$0.0488 & 0.179 & 2.66 \\
SchS34 & 0.9 & 1.19 & 2.44 & 0.909 & $-$0.206 & 0.175 & 2.68 \\
Sch$\gamma$1 & 0.891 & 1.2 & 1.49 & 0.884 & $-$0.103 & 0.182 & 2.63 \\
Sch$\gamma$2 & 0.911 & 1.11 & 0.978 & 0.954 & $-$0.028 & 0.154 & 2.75 \\
\hline
\end{tabular}
\caption{The nuclear mass fits corresponding to the models described
in the text. The values of $\sigma$ are given in MeV/fm$^{2}$ and
$n_0$ and $n_1$ are given in fm$^{-3}$.}
\end{table}

The mass fit is performed by minimizing $\Delta_{\mathrm{RMS}}$ for
all the experimentally measured mass excesses from
Ref.~\cite{Audi03}. For each nucleus, this involves computing the
neutron and proton densities using Eqs.~\ref{eq:n} and \ref{eq:np}, 
computing the bulk energy from the EOS of homoegenous matter
at these densities, then inserting this bulk energy into the nuclear
mass formula to compute the mass excess using Eq.~\ref{eq:massform}. 

As expected, there is a correlation between the surface symmetry
energy and the symmetry energy as shown by the increase in
$\sigma_{\delta}$ when going from model SchS28 to model SchS34.  This
is also the reason why NL4 gives a larger value of $\sigma_{\delta}$
than the other models. The value of $\sigma_{\delta}$ is nearly
unchanged by modifying $\gamma$, which changes the density dependence
of the symmetry. The values of $n_0$ in Table~\ref{tab:fit} are not
quite equal to the saturation density for homogeneous nuclear matter,
and this can be attributed to finite-size effects not captured in
Eq.~\ref{eq:n}. The other parameters are nearly unchanged between
models except for $n_1$ which is also sensitive to the symmetry
energy, as well as the Coulomb interaction.

\section{The Crusts}

In order to determine the composition and properties of the crust, the
energy at a fixed density as a function of the proton number and
atomic number of nuclei, and the number density of dripped neutrons,
$n_{n,drip}$, is minimized. The energy of matter in the neutron star
crust is given by
\begin{eqnarray}
\varepsilon(Z,A,n_{n,\mathrm{drip}}) &=& (n_n+n_p) \chi B(Z,A)/A +
\nonumber \\ 
&& (1-\chi) {\varepsilon}_{\mathrm{drip}}(n_{n,\mathrm{drip}})+
{\varepsilon}_{\mathrm{el}} (n_e)
\end{eqnarray}
This energy is minimized over the
three parameters $Z, A$, and $n_{n,\mathrm{drip}}$ at each density.
The volume fraction of matter inside nuclei, $\chi$, is determined from 
the relation
\begin{equation}
n_B = \chi (n_n+n_p) + n_{n,\mathrm{drip}} (1-\chi)
\end{equation}

\begin{figure}
\includegraphics[scale=0.42]{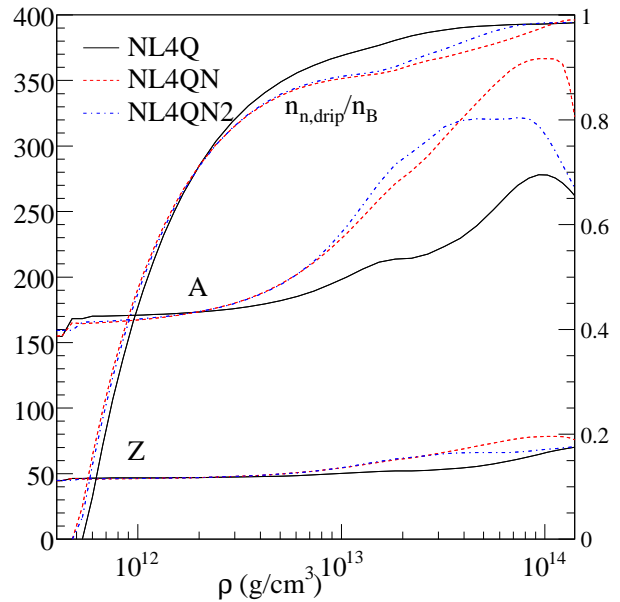}
\caption{A comparison of the composition of the neutron star crusts in
models NL4Q, NL4QN, and NL4QN2, designed to demonstrate the
sensitivity of the composition to the equation of state of low-density
neutron matter. The proton nuber, Z, the atomic number A (left axis)
and the number density of the dripped neutrons $n_{\mathrm{n,drip}}$ are displayed
(right axis). }
\label{fig:ld}
\end{figure}

The inner crust implied by the models NL4Q, NL4QN and NL4QN2 are
compared in Fig.~\ref{fig:ld}. While
the actual number density of dripped neutrons is not strongly modified
by modifying the neutron matter EOS, the nuclear size is modified by
50 \% or more. The larger energy cost of creating neutron matter in
with a more realistic neutron matter EOS is reflected in moving
neutrons into nuclei so as not to pay the energy cost. Because the
largest difference is between the models NL4Q and NL4QN2, it is clear
that most of the dependence on the low-density neutron matter EOS lies
at densities below 0.04 fm$^{-3}$. 

\begin{figure}
\includegraphics[scale=0.42]{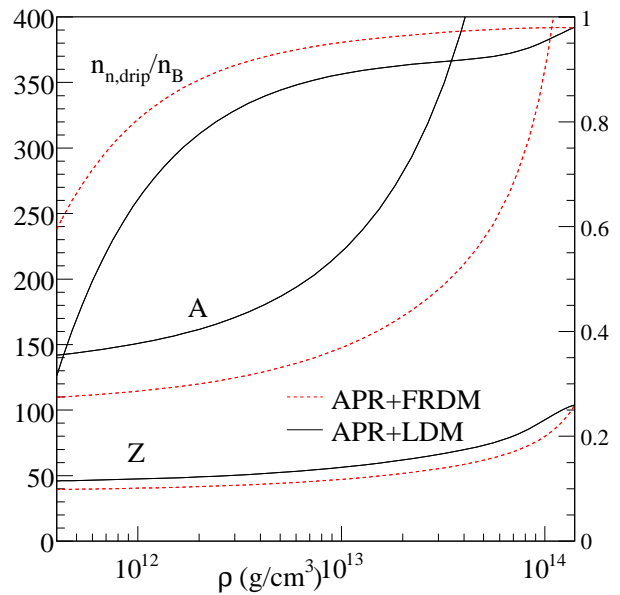}
\caption{A comparison the composition given the two different mass
models. The curve labeled LDM (solid lines) is obtained from
Eq.~\ref{eq:massform}, and the curve labeled FRDM (dashed lines) is
obtained from Ref.~\cite{Moller95}. The left axis is for $A$ and $Z$
and the right axis is for $n_{\mathrm{n,drip}}$.}
\label{fig:mass}
\end{figure}

In order to show the effect due to changing the mass model,
Fig.~\ref{fig:mass} shows the results for the APR EOS with the two
different models, the liquid drop model and the FRDM. The results are
qualitatively the same but quantitatively different. The form of the
mass model remains a significant uncertainty in the nature of the
neutron star crust, and is comparable to the other uncertainty
obtained from the symmetry energy as described below. 

\begin{figure}
\includegraphics[scale=0.42]{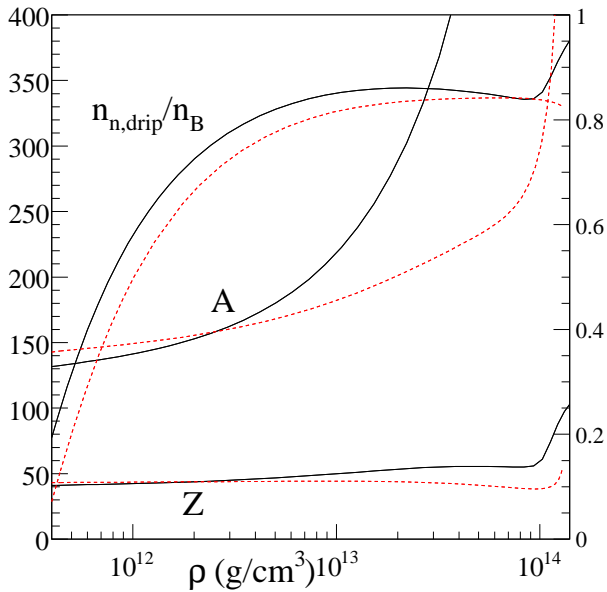}
\caption{A comparison of the present work (solid lines) with the
results of Ref.~\cite{Douchin01} (dashed lines) using the same input
equation of state, SLy4. The left axis is for $A$ and $Z$
and the right axis is for $n_{\mathrm{n,drip}}$.}
\label{fig:dh}
\end{figure}

\begin{figure}
\includegraphics[scale=0.42]{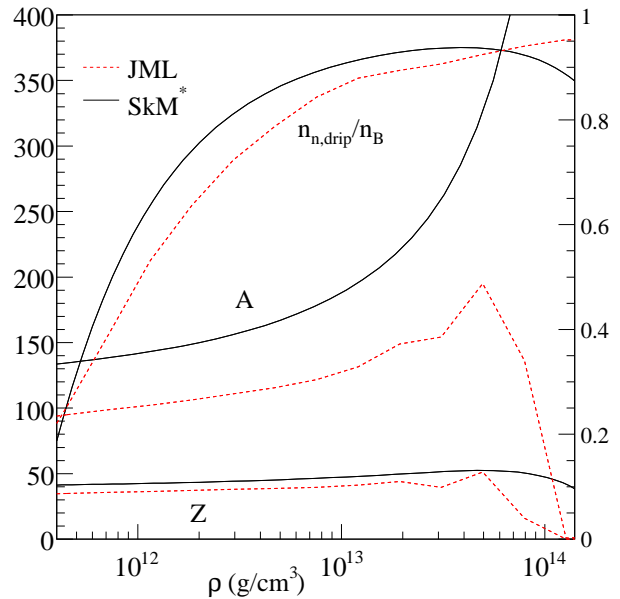}
\caption{A comparison of the present work (solid lines) with the
results from Ref.~\cite{Lattimer06} (labelled ``JML'' and plotted with
dashed lines) using the same input equation of state, Sk$M^{*}$.  The
sharp dropoff in JML at high-densities is due to the transition to
homogeneous nucleonic matter. The slightly jagged nature of the JML
results is due to the naive interpolation employed in this work and
is not necessarily present in the original table.
The left axis is for $A$ and $Z$ and the
right axis is for $n_{\mathrm{n,drip}}$. }
\label{fig:jl}
\end{figure}

Comparisons of the present model to that of
Refs.~\cite{Douchin01,Lattimer06} are given in Figs.~\ref{fig:dh} and
\ref{fig:jl}. The same input EOS for homogeneous matter as the
original reference is used in both cases. The results agree
qualitatively with the aforementioned works. The remaining differences
lie within the nuclear mass formula used, and they are within the
range of variation which is suggested by Fig.~\ref{fig:mass}. This
model (like all other models of the neutron star crust presently
available) cannot precisely predict the composition of the inner
crust. Nevertheless, it is qualitatively correct and is thus
useful for estimating the uncertainties due to the input EOS of
homogeneous matter.

\begin{figure}
\includegraphics[scale=0.42]{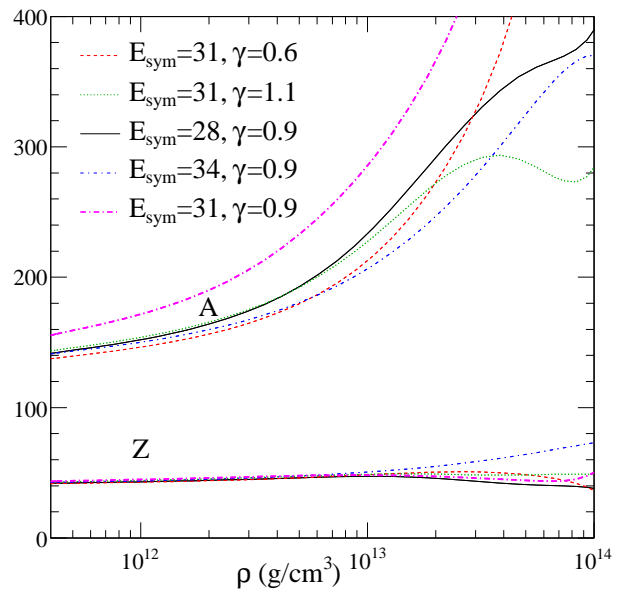}
\caption{A comparison of the composition of the crust given 
different symmetry energies. The bold solid line is the baseline
model, the dashed and dotted lines show the variation in
$\gamma$, and the dashed-dotted and thin solid line give the 
variation in the magnitude of the symmetry energy at the
saturation density.}
\label{fig:se}
\end{figure}

To compare the effect of the uncertainty in the symmetry energy,
Fig.~\ref{fig:se} shows the composition for the neutron star crust as
a function of density for the schematic equations of state with
different symmetry energies. The naive expectation is 
that a stronger symmetry energy tends to encourage nuclei to
become more isospin-symmetric. This is coupled, however, with the fact
that an increased symmetry energy will also raise the energy cost for
the dripped neutrons. These two effects together could force larger, more
symmetric nuclei, but this also affects the Coulomb and surface
energy contributions. The variation of the composition with the value
of the symmetry energy is not so clear, as the baseline model predicts
larger nuclei than either models with smaller or larger values of the
symmetry energy. 

\begin{figure}
\includegraphics[scale=0.42]{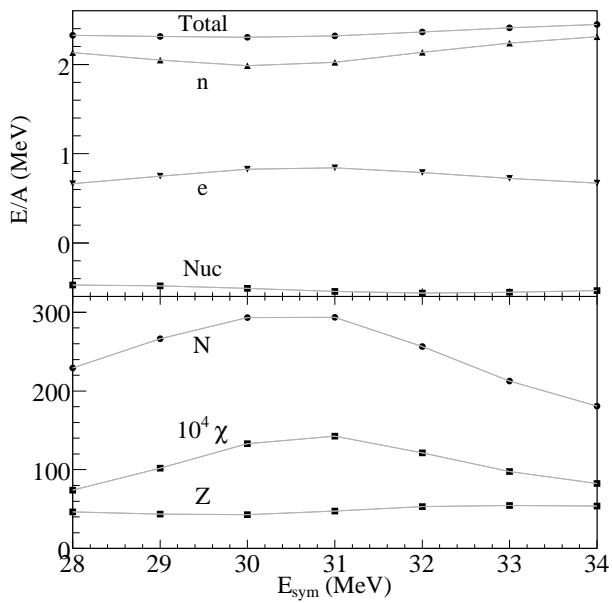}
\caption{A comparison of the composition of the crust given different
values of the symmetry energy at the saturation density at a fixed
density of $n_{B}=0.01$ fm$^{-3}$. The top panel gives the total
binding energy per baryon, and the separate contributions from dripped
neutrons (``n''), electrons (``e''), and nuclei (``Nuc''). The bottom
panel shows the neutron and proton number of nuclei as well as the
volume fraction, $\chi$.}
\label{fig:seden}
\end{figure}

In order to disentangle this result, more detailed results for
schematic models with different symmetry energies are given in
Fig.~\ref{fig:seden} at a fixed density of $n_{B}=0.01$ fm$^{-3}$.
Beginning with the larger symmetry energy (with a value at saturation
of 34 MeV) and proceeding downward, the expected result is obtained:
lower symmetry energies allow the system to create more
isospin-asymmetric nuclei. At low enough symmetry energies, however,
this becomes too costly as the electron contribution to the energy
increases (the proton number decreases, but the volume fraction
occupied by nuclei increases, thus the electron density must
increase). Instead, the system reponds by moving neutrons out of the
nuclei, which lowers the electron contribution, even though it
increases the contributions from nuclei and the dripped neutrons. This
is allowed, in part, because the nuclei are able to maintain a
relatively constant energy. They can do this because the surface and
Coulomb energy cost is cancelled by the bulk energy gain which results
from making nuclei with a larger (in absolute magnitude) bulk binding
energy. 

\begin{figure}
\includegraphics[scale=0.42]{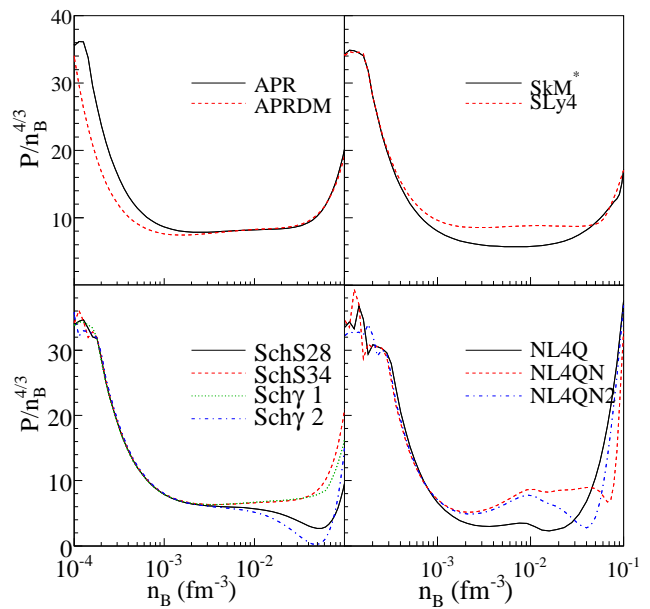}
\caption{A survey of the pressure of the crust as a function of
density, scaled by the baryon density, $n_B^{4/3}$. The curves are
nearly flat where the adiabatic index is expected to be nearly 4/3
since the baryon density nearly scales with the energy density. }
\label{fig:ep}
\end{figure}

Finally, Fig.~\ref{fig:ep} summarizes the pressure as a function
of the baryon density. The upper left panel shows the variation from
the different mass formulas, which is larger at lower densities. The
upper right panel shows the results for SkM$^{*}$ and SLy4. The
lower-left panel shows the variation allowed by the symmetry energy,
and variations of up to a factor of two in the pressure are implied by
the uncertainty in the symmetry energy. The pressure appears sensitive
to the magnitude of the symmetry energy and its dependence on
density. Finally, the lower-right panel shows the pressure for the
NL4Q-related models. Note that for model NL4Q, the anomalously small
EOS of low-density neutron matter underestimates the pressure in the
inner crust. All of the models are connected (sometimes
discontinuously) to the EOS from Ref.~\cite{Baym71} at low densities
thus giving the scatter in the pressure at the lowest densities given
in this figure. Particularly interesting is that the pressure at the
crust at the higher densities is nearly independent of the mass model
(as shown in the upper left panel), which may indicate that
astrophysical observables which are sensitive to the pressure of the
crust rather than the composition are good probes of the nuclear
symmetry energy, assuming that the description of low-density neutron
matter is correct.

The transition density to homogeneous nuclear matter is computed 
by noting the density at which the energy per baryon of nuclear matter
becomes smaller than that of the heterogeneous phase. The results for
the transition density are given in Table~III.  The transition
densities from this work are given in the second column and third
column contains the transition densities for previous works with
similar input EOSs for comparison.  These transition densities may be
underestimates, as the ability of nuclei to deform slightly will
decrease the energy of the heterogeneous phase and thus increase the
transition density. This will be addressed in further work. Note that,
as in the composition discussed above, the transition densities depend
non-trivially on the symmetry energy.

\begin{table}
\label{tab:nt}
\begin{tabular}{ccc}
Model & $n_{t}$ fm$^{-3}$ & \\
\hline
APR & 0.0522 & \\
SkM$^{*}$ & 0.0434 & 0.045~\cite{Lattimer06} \\
SLy4 & 0.0669 & 0.076~\cite{Douchin01} \\
NL4Q & 0.0344 & \\
NL4QN & 0.0409 & \\
NL4QN2 & 0.0333 & \\
Sch & 0.0584 & \\
SchK210 & 0.0585 & \\
SchK250 & 0.0591 & \\
SchS28 & 0.0368 & \\
SchS34 & 0.0416 & \\
Sch$\gamma$1 & 0.0676 & \\
Sch$\gamma$2 & 0.0641 & \\
\hline
\end{tabular}
\caption{Caption here.}
\end{table}

\section{Conclusions}

The composition of the neutron star crust is still partially unknown,
due to uncertainties in the nuclear mass formula and the equation of
state. The composition (and to a lesser extent, the overall pressure)
is quite sensitive to the equation of state of low-density neutron
matter, and the nuclear symmetry energy, both its magnitude and its
density dependence. The dependence of the composition on the symmetry
energy is not monotonic, as models with moderate symmetry energies can
have larger nuclei than models with lower or higher symmetry
energies. To the extent to which neutron stars depend on the
composition, this means that it is important to explore the full range
of variation in the crust allowed by the present knowledge of the
input nuclear physics, while ensuring that the EOS is constrained by
what is already known about the EOS of low-density neutron
matter. Nuclear experiments will continue to provide better
constraints on the symmetry energy, including from the PREX
experiment~\cite{Michaels00,Horowitz01b} to measure the neutron skin
thickness of lead at Jefferson Lab and from intermediate-energy
heavy-ion collisions as has been done in Ref.~\cite{Tsang04}.

It remains to be seen if these results persist in the more microscopic
models which include pairing, corrections beyond the Wigner-Seitz
approximation, long-range correlations, and better treatments of the
nuclear structure.

\section{Acknowledgements}

The author would like to thank Ed Brown, Joe Carlson, Alex Gezerlis,
Jim Lattimer, Bill Lynch, Sanjay Reddy, Sergio Souza, and Anna Watts
for useful discussions related to this work. This work is supported by
Joint Institute for Nuclear Astrophysics at MSU under NSF-PFC grant
PHY 02-16783.

\bibliography{paper}
\bibliographystyle{apsrev}
\end{document}